# Free water DTI estimates from single b-value data might seem plausible but must be interpreted with care

Marc Golub[1], Rafael Neto Henriques[2] †, Rita G. Nunes[1] †

[1]*Institute for Systems and Robotics and Department of Bioengineering, Instituto Superior Técnico, Universidade de Lisboa, Lisbon, Portugal*
[2]*Champalimaud Research, Champalimaud Centre for the Unknown, Lisbon, Portugal*

†These authors contributed equally to this work.

Corresponding Author: Rita G. Nunes

        Institute for Systems and Robotics,

        Instituto Superior Técnico,

        Av. Rovisco Pais, 1

        1049-001 Lisboa, Portugal

        Phone: +351.218418277

        Fax: +351.218418291

        E-mail: ritagnunes@tecnico.ulisboa.pt

# Abstract

Purpose
Free water elimination diffusion tensor imaging (FWE-DTI) has been widely used to distinguish increases of free water (FW) partial volume effects from tissue's diffusion in healthy ageing and degenerative diseases. Since the FWE-DTI fitting is only well posed for multi-shell acquisitions, a regularized gradient descent (RGD) method was proposed to enable application to single-shell data, more common in the clinic. However, the validity of the RGD method has been poorly assessed. This study aims to quantify the specificity of FWE-DTI procedures on single- and multi-shell data.

Methods
Different FWE-DTI fitting procedures were tested on an open-source in vivo diffusion dataset and single- and multi-shell synthetic signals, including the RGD and standard non-linear least squares (NLS) methods. Single-voxel simulations were carried out to compare initialization approaches. A multi-voxel phantom simulation was performed to evaluate the impact of spatial regularization when comparing between methods. To test the algorithms' specificity, phantoms with two different types of lesions were simulated: with increased mean diffusivity (MD) or with increased FW.

Results
Plausible parameter maps were obtained with RGD from single-shell in vivo data. The plausibility of these maps was shown to be determined by the initialization. Tests with simulated lesions inserted into the in vivo data revealed that the RGD approach cannot distinguish FW from tissue MD alterations, contrarily to the NLS algorithm.

Conclusion
RGD FWE-DTI has limited specificity and, thus, its results from single-shell data should be carefully interpreted. When possible, multi-shell acquisitions and the NLS approach should be preferred instead.

# 1 Introduction

Diffusion-weighted Magnetic Resonance Imaging (dwMRI) is a non-invasive imaging modality sensitive to microscopic tissue properties beyond the macroscopic spatial resolution offered by current MRI scanners (1). The information captured by dwMRI is multi-dimensional and its sensitivity to different diffusion properties depends on the acquisition parameters. For example, diffusion tensor imaging (DTI) can estimate diffusion anisotropy from dwMRI images acquired with different gradient directions for a single level of diffusion-weighting (i.e. single-shell acquisition) (2). Another technique, diffusion kurtosis imaging, allows estimating non-Gaussian diffusion properties from dwMRI images acquired with different contrast levels (i.e. multi-shell acquisition) (3,4). Although both are sensitive to tissue alterations undetected by conventional structural images (1,5), the derived metrics are not specific to concrete microstructural properties and are thus hard to interpret (6–8).

To improve dwMRI's specificity, microstructural models have been introduced to directly extract biophysical measures (6,9–11). However, assumptions and constraints are often required to minimize the number of parameters to be estimated (8,12). For instance, several studies used a simplified tissue model, with two compartments assigned to intra- and extra-cellular media to estimate axonal water fraction and diameter (13–17). Two-compartmental models have also been used to decouple free water (FW) contributions from the tissue's diffusion tensor in FW elimination DTI (FWE-DTI) (18,19).

FWE-DTI has attracted a huge interest to characterize diffusion alterations in the context of healthy aging (20,21) and neurodegeneration associated with Alzheimer's Disease (22), Parkinson's Disease (23), and Traumatic Brain injury (24). These studies suggest that FWE-DTI is fundamental to decouple changes associated with microstructural alterations from increases in partial volume near interfaces between tissue and cerebrospinal fluid (CSF); in particular, partial volume increases associated with gross tissue atrophy and ventricular enlargement. Moreover, compared to suppression techniques based on fluid attenuated inversion recovery (FLAIR) (25,26), FWE-DTI has the advantage of providing a FW fraction estimate, a potential surrogate marker for edema (19,27).

As any other two-compartmental model, FWE-DTI is only well-posed for multi-shell acquisitions (28,29). Previous studies successfully showed that the apparent FW fraction can be estimated from such acquisitions by combining linear and non-linear approaches (25,29,30). However, since single-shell acquisitions are clinically more prevalent due to time constraints, several single-shell FWE-DTI procedures have been proposed based on spatially regularized gradient descent (RGD) algorithms, requiring careful initialization (19,27,31). Although single-shell FWE-DTI procedures were shown to provide plausible FW maps, their use in clinical applications remains controversial - these algorithms have been poorly validated (plausibility does not imply specificity) and no clear theoretical explanation has been provided on how they recover information on the bi-exponential nature of the FWE-DTI model. Moreover, no single-shell FWE-DTI procedure is currently available in an open-source software, hampering independent and objective validation and comparison with other FW estimation techniques.

In this study, state-of-the-art FWE-DTI fitting procedures RGD algorithms are revisited. Our main objective was to quantify the specificity of FWE-DTI estimates when extracted from single- and multi-shell dwMRI acquisitions. The plausibility of FWE-DTI estimates is first illustrated on openly available dwMRI data. Based on numerical simulations, we investigated which FWE-DTI fitting procedures steps determine the plausible contrast of obtained FW maps. At last, the specificity of FW maps from single- and multi-shell acquisitions was quantitatively assessed on synthetic phantoms to which different lesions were introduced. For reproducibility, all tested FWE-DTI fitting procedures were incorporated into the open-source library diffusion in python (DIPY) (32).

## 2 Theory

### 2.1 FWE-DTI model

In FWE-DTI, a two-compartmental model is used to account for FW contamination (18,19). Assuming that the repetition time (TR) is sufficiently long to ignore T1 relaxation, this model can be described by:

$$S_k(D_t, v, \rho_t, \rho_w, T2_t, T2_w) = v\rho_t \exp\left(-\frac{TE}{T2_t}\right)\exp(-b_k n_k^T D_t n_k) + (1-v)\rho_w \exp\left(-\frac{TE}{T2_w}\right)\exp(-b_k D_w),$$

[1]

where $v$ is the volumetric tissue fraction, $\rho_t$ and $\rho_w$ are the tissue and water proton densities, $T2_t$ and $T2_w$ are the tissue and FW transverse relaxation times, TE is the acquisition echo time, $D_t$ is the tissue's apparent diffusion tensor (with 6 independent elements (2)), $D_w = 3\ \mu m^2 ms^{-1}$ is the constant diffusivity of isotropic FW at body temperature ($\cong 37°$) (33), $b_k$ is the diffusion weighting for the $k^{th}$ acquisition and $n_k$ the respective gradient direction (normalized column vector; $T$ denotes the transpose). This model assumes no water exchange between the two compartments and that both display Gaussian diffusion (19).

Since decoupling the volumetric tissue fraction $v$ from parameters $\rho_t, \rho_w, T2_t$, and $T2_w$ would require additional relaxometry measurements, the effective tissue water fraction is often defined as $f = vS_t/S_0$, where $S_0$ is the signal measured at b-value $b_0 = 0$ (i.e. $S_0 = vS_t + (1 - v)S_w$), $S_t$ is the reference $b_0$ signal from voxels containing only grey matter (GM) and white matter (WM) tissues (assuming all tissue presents a single $T2_t$ value, $S_t = \rho_t \exp(-TE/T2_t)$), and $S_w$ is the reference $b_0$ signal from voxels containing only FW ($S_w = \rho_w \exp(-TE/T2_w)$). Inserting the effective tissue water fraction $f$ into Eq. 1, the FWE-DTI model can be rewritten as:

$$A_k(D_t, f) = f \exp(-b_k n_k^T D_t n_k) + (1 - f) \exp(-b_k D_w),$$

[2]

where $A_k$ is the diffusion signal attenuation (i.e., $A_k = S_k/S_0$). From Eq. 2, the effective FW fraction is defined as $f_w = 1 - f$. Note that if $\rho_t, \rho_w, T2_t$, and $T2_w$ are known, the effective fractions can be converted to actual volumetric fractions $v = fS_w/(S_t + f(S_w - S_t))$.

Despite its simplicity, Eq. 2 is a two-compartmental model and has thus a flat fitting solution landscape, i.e. similar residuals are observed for different $D_t$ samples for any given $f \in [0,1]$ as shown in (28) so that choosing the most viable pair (Dt, f) is not straightforward.

## 2.2 Regularized Gradient Descent fitting procedure

To fit the FWE-DTI model to single-shell data, previous studies proposed using regularized gradient descent (RGD) algorithms with careful parameter initializations (19,27,31,34). The different initialization procedures and the RGD algorithm implemented and tested in this study are described in the next subsections. Notably, although these procedures were designed for single-shell acquisitions, all these can also be adapted to multi-shell datasets (vide infra).

### 2.2.1 Parameter initialization

Three different strategies were tested to initialize $f$ in the RGD algorithm.

Initialization based on the T2-weighted images: this empirical estimation (here referred to as $f_{S0}$) assumes that regions with higher FW fraction present hyperintense T2-weighted signals ($S_0$) (19):

$$f_{S0} = 1 - \log(S_0/S_t^r)/\log(S_w^r/S_t^r),$$

[3]

where $S_t^r$ and $S_w^r$ are, respectively, the reference signal intensities for voxels containing either tissue or free diffusing water only, and can be estimated from regions of interest placed on deep WM or the ventricles (19). To avoid implausible estimates, $f_{S0}$ values are constrained:

$$\frac{\min(\hat{A}_k - \exp(-b_k D_w))}{\max(\exp(-b_k \lambda_{min}) - \exp(-b_{kDw}))} \leq f_{S0} \leq \frac{\max(\hat{A}_k - \exp(-b_k D_w))}{\min(\exp(-b_k \lambda_{max}) - \exp(-b_{kDw}))},$$

[4]

where $\hat{A}_k$ is the measured signal decay, λmin and λmax are the minimum and maximum expected tissue diffusivities (0.1 and 2.5 $\mu m^2 ms^{-1}$, respectively (19)). The equation was modified for compatibility with multi-shell data, and to correct for the swapped denominators in (19).

Initialization based on a tissue's Mean Diffusivity prior: to avoid relying on the non-quantitative $S_0$ images, tissue water estimates (here referred to as $f_{MD}$) can be initialized based on a fixed prior for the tissue's Mean Diffusivity (MD) (31):

$$f_{MD} = \frac{\exp(-b\,MD) - \exp(-bD_w)}{\exp(-bMD_t^{ref}) - \exp(-bD_w)},$$

[5]

where $MD$ is computed with standard DTI, $MD_t^{ref}$ is the fixed tissue's MD prior (set to 0.6 $\mu m^2 ms^{-1}$ (31,34)), and $b$ the single shell b-value. For multiple-shell acquisitions, one b-value is empirically selected (1 $ms\ \mu m^{-2}$ in our study) which provides sufficient diffusion contrast while minimizing non-Gaussian diffusion effects not considered by FWE-DTI (19). Notably, voxels with higher MD values will present higher $f_{MD}$. Unlike $f_{S0}$, $f_{MD}$ is constrained by 0 and 1.

Hybrid initialization ($f_{hybrid}$): to combine T2-weighted and dwMRI information, a log-linear interpolation can be performed between $f_{S0}$ and $f_{MD}$ estimates (31,34):

$$f_{hybrid} = f_{S0}^{1-\alpha} \times f_{MD}^{\alpha},$$

[6]

where $\alpha$ determines their relative weights. To assign a higher weight to $f_{MD}$ in regions with T2-weighted signals closer to typical tissue intensities (i.e. healthy tissue) and higher weights to $f_{S0}$ in hyperintense voxels (i.e. edematous tissue), $\alpha$ is set to the initial tissue water fraction computed by Eq. 3 but unconstrained by $f_{min}$ and $f_{max}$ (Eq. 4) (34).

Initialization of tissue's diffusion tensor: The tissue fraction initializations were used to estimate the normalized tissue's signal attenuation $[A_t]_k = (\hat{A}_k - (1 - f) \exp(-b_k D_w))/f$. For each initialization ($f_{S0}$, $f_{MD}$, and $f_{hybrid}$), an initial estimate of $D_t$ was obtained by fitting the standard DTI model to $[A_t]_k$.

### 2.2.2 Regularized Gradient Descent

Here, we implemented an adapted version of the FWE-DTI fitting framework that constrains $D_t$ to be spatially smooth by introducing a regularization term into the minimization functional (19):

$$L(D_t, f) = \int_\Omega \left[ \sum_{k=1} \left( A_k(D_t, f) - \hat{A}_k \right)^2 + \omega \sqrt{|\gamma(D_t)|} \right] d\Omega,$$

[7]

where $\Omega$ represents the image domain, $A_k(D_t, f)$ is the predicted signal attenuation given by Eq. 2, $\hat{A}_k$ is the measured signal attenuation for the $k^{th}$ gradient direction, $|\gamma(D_t)|$ is the determinant of the induced metric which acts as a regularizer, and $\omega$ a hyperparameter for controlling its weight (19). The metric tensor is computed from the spatial derivatives of $D_t$ and was chosen to be Euclidean to reduce computational burden (27). These concepts are borrowed from the field of differential geometry; application to diffusion tensors is further explained in (19,35); the Euclidean metric implementation is described in (27).

Minimization of Eq. 7 can be done with a gradient descent scheme, following the iteration rules provided in (19,27). Here, a small correction is proposed to the fidelity term $\Delta F_i$ in (19), obtained by differentiating the first term of Eq. 7 with respect to each independent diffusion component $X^i$ ($i$ ranges from 1 to 6):

$$\Delta F^i = \sum_{k=1}(A_k(D_t, f) - \hat{A}_k) f \exp(-b_k n_k^T D_t n_k)\left(b_k n_k^T \frac{\partial D_t}{\partial X^i} n_k\right),$$

[8]

where $\frac{\partial D_t}{\partial X^i}$ is the partial derivative of $D_t$ with respect to $X^i$. Minimization of the second term in Eq. 7 gives rise to the Laplace-Beltrami operator (19,27).

Differentiating Eq. 7 with respect to $f$ gives the tissue fraction increment, again slightly different from that presented in (19):

$$\Delta f = \sum_{k=1}(A_k(D_t, f) - \hat{A}_k)(f \exp(-b_k n_k^T D_t n_k) - \exp(-b_k D_w)).$$

[9]

### 2.2.3 Dealing with voxels containing only free water

FWE-DTI estimates are not well defined for voxels containing only FW (25,29). Since their signal is well-described by a single exponential with high isotropic diffusion (i.e. $\sim 3\ \mu m^2 ms^{-1}$), $D_t$ can erroneously be fitted with a diffusion tensor with a large trace and FW assigned any value between 0 and 1. To ensure a FW estimate close to 1, initial parameter estimates with high tissue MD can be re-adjusted to have $f = 0$ and $D_t$ assigned null elements. Tissue MD is classified as high if above $1.5\ \mu m^2 ms^{-1}$ (25,29). To avoid low precision FWE-DTI estimates in regions with low tissue contributions, refined tissue MD and fractional anisotropy (FA) estimates are set to zero when the refined $f$ is below 0.1 (25,29).

## 2.3 Standard least squares fitting procedure

The FWE-DTI model has a unique $(D_t, f)$ solution when multi-shell data is available, providing a reference for regularized single-shell fitting procedures. Here a 2-step minimization based on a combination of standard weighted linear least squares (WLS) and non-linear least square (NLS) routines was used as reference multi-shell FWE-DTI fitting procedure (25,29).

# 3 Methods

## 3.1 MRI Data

A dwMRI dataset of a healthy volunteer acquired in a Siemens Prisma 3T scanner was used (https://digital.lib.washington.edu/researchworks/handle/1773/33311), including b-values $0.2, 0.4\ ms\ \mu m^{-2}$ (8 directions per shell, sampled twice) and $1, 2, 3\ ms\ \mu m^{-2}$ (90 directions per shell); an unweighted image was acquired every 8th or 9th volume. Other relevant imaging parameters: isotropic resolution of $2\ mm$, multi-band factor of 3, $TR = 3000\ ms$, $TE = 74\ ms$ and flip angle of $72\ °$. This dataset had been previously pre-processed to correct for eddy currents distortions and motion using FSL tools (36,37) and incorporated into DIPY. In addition, here this dataset is corrected for B1 field inhomogeneities using "dwibiascorrect" from MRTrix3 with the FSL-FAST option (38).

This data was used to provide a qualitative assessment of the different FWE-DTI estimates before our quantitative analysis. For this, all dwMRI images acquired with b-values larger than $1\ ms\ \mu m^{-2}$ were removed to avoid non-Gaussian diffusion effects (25,29). Diffusion parameter maps were extracted and compared between: 1) standard DTI model (using DIPY's WLLS fitting (32)); 2) FWE-DTI model using the regularized gradient descent algorithm including only the data with $b = 0$ and $1\ ms\ \mu m^2$ (RGD, single-shell); 3) FWE-DTI model using the regularized gradient descent algorithm on all b-values $\leq 1\ ms\ \mu m^2$ data (RGD, multi-shell); 4) FWE-DTI model using the standard non-linear approach (25,29) (NLS, multi-shell). For simplicity, the single- and multi-shell RGD algorithm for this first assessment was initialized with the hybrid initialization technique (34).

## 3.2 Simulations

Quantitative analyses to assess the robustness of different FWE-DTI fitting steps were first performed based on single- and multi-voxel synthetic phantoms. For this purpose, dwMRI signals were numerically generated using Eq.1 with: $\rho_t$ and $\rho_w$ set to typical proton density values (70% and 100% (39); TE set to $74\ ms$ (as for the in vivo dataset); $T2_t$ and $T2_w$ set to $80\ ms$ and $500\ ms$, typical 3T transverse relaxation times for tissue (40) and CSF (41). Ground truth tensors were simulated considering reference values for the $D_t$'s eigenvalues: $\lambda_1^{WM} = 1.6$, $\lambda_2^{WM} = 1.5$ and $\lambda_3^{WM} = 0.3\ \mu m^2 ms^{-1}$ (typical for WM (29)), corresponding to $FA_t^{WM} = 0.7$ and $MD_t^{WM} = 0.8\ \mu m^2 ms^{-1}$. Synthetic signals were simulated using the forward model (Eq. 1) and Rician noise was added to achieve a signal-to-noise ratio (SNR) of 40 (reference $S_0$ SNR for FW voxels).

<u>Single-voxel simulations:</u> to compare between initialization methods, single-voxel simulations were repeated for $D_t$ with different $MD_t$ ground truth values (sampled between 0.1 and $1.6\ \mu m^2 ms^{-1}$). To maintain $FA_t^{WM}$, the eigenvalues of $D_t$ were computed by $\lambda_1^{gt} = c\lambda_1^{WM}$, $\lambda_2^{gt} = c\lambda_2^{WM}$, and $\lambda_3^{gt} = c\lambda_3^{WM}$, where $c = MD_t/MD_t^{WM}$. Simulations were also repeated for different tissue water fraction values $f$ linearly spaced between 0 and 1, converted to volume fractions $v$ to generate the synthetic dwMRI signals using Eq.1. For each ground truth $(f, MD_t)$ pair, 100 different directions were considered for the principal orientation of $D_t$ and for each orientation, single-shell signals were generated along 32 gradient directions with $b = 1\ \mu m^2 ms^{-1}$. and six $S_0$ images. Synthetic Rician noise was added for 100 noise instances and the initialization methods were applied to each single-voxel signal. The median and inter-quartile ranges of the $f$, $FA_t$ and $MD_t$ estimates were computed over the 100 repeated $D_t$ directions times 100 noise instances.

<u>Multi-voxel phantom:</u> this phantom was designed to assess if applying spatial regularization to the RGD algorithm improves the estimates of FWE-DTI in a best-case scenario (i.e. phantom with a smooth $D_t$ field). For this purpose, a multi-voxel phantom (21×21×21 voxels) of a cylindrical fiber (radius of 7 voxels) was designed, with flat ground truth $D_t$. The fiber was contaminated with three levels of effective FW fraction $f_w = 0.1, 0.4$ and $0.7$ that increased along the radial direction while kept constant along the axial direction. To compare the FWE-DTI with DIPY's multi-shell fitting procedure (NLS FWE-DTI (25)), besides the single-shell scenario, simulations considered a multi-shell dataset with b-values of 0.5 and $1\ ms\ \mu m^2$ (32 directions per shell), with six $S_0$ images. Standard DTI and the RGD procedure for the FWE-DTI model were applied to both data (RGD single- and multi-shell for short), while the NLS FWE-DTI was applied to the multi-shell data only. The estimated $f_w$, $FA_t$ and $MD_t$ maps were visually assessed, and the medians of the initial and final FWE-DTI parameter estimates compared to the ground truth values.

In-vivo data with simulated lesions: the aim was to evaluate the specificity of FWE-DTI parameter estimates, i.e. whether these enable decoupling alterations to $D_t$ from changes in the degree of FW contamination. For this, two types of synthetic lesions were inserted into a representative dwMRI brain dataset: 1) lesions with increased FW content (FW lesion); and 2) lesions with increased $MD_t$ (MD lesion). Ground truth FWE-DTI parameter maps were obtained by applying the gold standard NLS FWE-DTI technique (25). Lesions covering a spherical volume radius of 14 $mm$ ) were placed in a WM region near the superior portion of the left internal capsule. The lesions were generated by increasing the $f_w$ ground truth values to 0.6 (FW lesion) or by increasing $MD_t$ to 1.1 $\mu m^2 ms^{-1}$ without changing the ground truth $FA_t$ (as described for the single-voxel simulation). The GT parameters were plugged into (Eq.1) to generate, for each lesion type, single- and multi-shell data with the same number of directions and b-values as described for the multi-voxel phantom. All non-diffusion parameters of the lesions were as for the synthetic phantoms. Standard DTI and RGD FWE-DTI (for the best performing initialization technique according to the phantom simulations) were applied to all single- and multi-shell datasets, while the NLS algorithm was applied only to the multi-shell data. The multi-shell data were also processed with a modified NLS algorithm, using the best performing single-shell initialization for a fairer comparison (same initialization). The estimated scalar maps were compared with the ground truth parameters to assess specificity.

For every run of the RGD routine, unless stated otherwise, the number of iterations was 200, the learning rate was 0.0005, and the spatial regularization operator was turned off halfway ($\omega$ was set to 0 at iteration 100) according to (19). To promote the reproducibility of our results, the code used for all simulations are available on an open source repository (https://github.com/mvgolub/FW-DTI-Beltrami).

# 4  Results

Figure 1 shows the estimated scalars maps from a healthy human brain. The $f_w$ maps show values near one for all FWE-DTI fitting procedures in regions comprising the brain ventricles and surrounding the parenchyma (first row). The corresponding $MD_t$ values were always lower than obtained using standard DTI (second row). Both RGD single- and multi-shell procedures provided $MD_t$ maps with lower grey to WM contrast compared to NLS FWE-DTI. The FA maps look similar for all methods (third row), but the FWE-DTI estimates were higher than the DTI estimate, particularly for WM regions.

The results obtained for the single-voxel simulations are presented in Figure 2. For the case of fixed ground truth $MD_t$ =0.6 $\mu m^2 ms^{-1}$ and varying $f_w$ (first column), the initialization based on $MD_t$ (blue markers) shows the smallest deviations to the ground truth line (in orange). When $f_w$ was fixed at 0.2, 0.5 or 0.8 and $MD_t$ deviated from 0.6 $\mu m^2 ms^{-1}$ (second, third and fourth columns), the $MD_t$ and hybrid (green) initializations differs more from the ground truth values. The performance of the $S_0$ based initialization (red markers) was invariant to the ground truth $MD_t$ (second, third and fourth columns), but the FWE-DTI estimates present a constant bias.

Scalar maps estimated for the single-shell dataset of the synthetic multi-voxel phantom are presented on the left side of Figure 3. Although no spatial variation was simulated on $MD_t$ and FA ground truth maps (first column), $f_w$ contamination induced a spatial variation on the standard DTI FA and MD maps (second column). FA, and $MD_t$ initial estimates obtained using the hybrid method (third column) show a lower spatial dependence compared to DTI estimates. The fourth column shows the FWE-DTI estimates refined using the RGD single-shell algorithm, showing identical contrast to the initial estimates. The median and interquartile ranges of the FWE-DTI parameter errors computed before ('init') and after ('est') applying the RGD algorithm are shown on the right side of Figure 3 for all initialization methods (left to right: $S_0$, $MD_t$ and hybrid methods). The latter plots confirm that refined estimates present similar accuracy and precision to their initial estimates, with the hybrid initialization resulting in smaller bias levels.

Figure 4 shows the analogous results for the multi-shell dataset of the multi-voxel phantom. Both hybrid initialization and RGD estimates present higher precision (lower interquartile ranges) than the NLS estimates, but they show lower estimate precision (higher deviation between estimates median and ground truth values). The interquartile range for the refined RGD estimates ('est') are slightly lower than the initial estimates ('ini') (right side of Figure 4).

The results for the synthetic lesions are presented in Figures 5 and 6 for FW and MD lesions, respectively. All FWE-DTI fitting routines were able to estimate the high $f_w$ increases in the FW lesion area (true positives pointed by the cyan arrows). Both the standard DTI, and single-shell RGD erroneously estimated an increased MD (false positives pointed by red arrow) in the FW lesion. Moreover, this FWE-DTI technique removed the $MD_t$ contrast between GM and WM regions. For MD lesions (Figure 6), the single- and multi-shell RGD algorithms (third and fourth columns) erroneously estimated increased $f_w$ values in the lesion (false positives pointed by the red arrow). The slight increase in MD (second row) is, however, still captured by the RGD particularly for the multi-shell dataset. Both NLS runs detected increased MD in the lesion without overestimating FW or impacting FA. NLS was specific to both lesion types even when initialized with the single-shell hybrid method estimates (last column of Figures 5 and 6).

These results are quantitatively summarized in Figure 7, where the median and interquartile ranges for estimates inside the lesion mask are shown for the initial hybrid estimation, RGD single/multi-shell and NLS* fits (red, blue, green and cyan, respectively). Deviation of RGD quantities from the ground truth median (grey line) and interquartile ranges (grey shadow area) are smaller for the FW lesion (first column) than for the MD lesion (second column). NLS produced increased interquartile ranges (a consequence of its lower precision); however, it produces median values that closely match the expected ground truth values which confirming its higher accuracy.

# 5 Discussion

The FWE-DTI model was designed to quantify the fraction of free diffusing water molecules in biological tissue (18,19). Despite being based on a simplistic two-compartmental model, several studies showed FWE-DTI can be useful to eliminate confounding FW partial volume effects from standard DTI metrics, particularly in studies including subjects with varying degrees of tissue maturation or atrophy (20–22,30,31). The application of FWE-DTI to single-shell acquisitions remains, however, controversial. While some studies showed that plausible FW fractions can be obtained using RGD algorithms (19,27,34)], others have argued that such maps are unreliable since this model is degenerate for single-shell acquisitions (25,29). We aimed to address this controversy using representative in vivo data of a healthy human brain and synthetic signals with known ground truth parameters.

## 5.1 FWE-DTI provides plausible estimates

We confirm that plausible maps can be obtained from FWE-DTI applying RGD algorithms to single-shell datasets; FW fraction estimates obtained in this way present the expected hyperintensities in the cerebral ventricles and subarachnoid space and hypo-intensities in deep

WM (Figure 1). Moreover, RGD estimates fitted to a healthy subject single-shell dataset provides similar contrasts to well-posed standard FWE-DTI procedures fitted to multi-shell datasets. However, we stress that plausibility does not imply specificity (vide infra).

## 5.2 Initialization determines FWE-DTI plausibility

Single-shell FWE-DTI fitting procedures are based on two main steps: 1) fast parameter initialization; 2) refinement of initial estimates using RGD algorithms.

Considering the theory behind the initialization methods, a common feature is that all resort to prior information. Particularly, initialization based on T2-weighted information uses priors on the typical pure FW and tissue signals, initialization based on MD assumes a constant prior for $MD_t$, while the hybrid initialization is just a log interpolation between the former techniques. Resorting to these priors, a well-posed solution for an initial FW fraction (one unknown) can be obtained assuming that all other parameters are known - the water and tissue signals acquired with b-value=0 for the T2-weighted initialization (Eq. 3), or $MD_t$ for the MD initialization (Eq. 5).

Refining the estimates when using the RGD algorithm attempted to use spatial information to improve the accuracy and precision of FWE-DTI estimates. Our numerical single-voxel simulations revealed that the initialization methods provide FW estimates somewhat sensitive to changes to the ground truth FW (Figure 2). Additionally, our synthetic multi-voxel phantom simulation showed that the FWE-DTI estimates refined by the RGD algorithm match the initial estimates when only single-shell data is provided (Figure 3). Based on these findings, we show that the plausibility of FWE-DTI initializations is the main determinant factor for the plausibility of FWE-DTI single-shell estimates.

## 5.3 Comparison across FWE-DTI initialization methods

Since initialization methods determine the plausibility of FWE-DTI contrasts for single-shell data, it is crucial to compare their robustness. Here, three initialization strategies were explored. We show that FWE-DTI initialization matches its ground truth only under certain conditions. Particularly, the $f_w$ estimates for MD initialization matches the ground truth identity line only when the MD prior is identical to the ground truth values of 0.6 $\mu m^2 ms^{-1}$ (first panel of Figure 2), while $MD_t$ estimates for $S_o$ initialization approach the ground truth identity line for synthetic signals generated with low FW (sixth panel of Figure 2).

In previous studies by Ismail and colleagues (31,34), these initialization methods were also compared, reporting final estimates closely matching the simulation ground truth values. However, the biases that can be induced when ground truth parameters deviate from the assumed priors had been barely considered. Indeed, our study shows that signals generated with $MD_t$ larger than 0.6 $\mu m^2 ms^{-1}$ can substantially inflate the FW fraction estimates for the MD initialization, leading to underestimated $MD_t$ and overestimated FA values. On the other hand, although less dependent on the ground truth $MD_t$, the FW fraction estimates based on $S_0$ images present biases that depend on the ground truth FW content. Regarding the hybrid method, its estimates present an intermediate behavior, consistent with the fact that it interpolates between the MD and the $S_0$ based estimates, resulting in biases that depend on both $MD_t$ and FW ground truth values. As this method presented lower biases than the MD based initialization for the single-voxel simulations and the lowest errors for the synthetic phantom (Figure 3), this technique was used for producing Figures 5, 6 and 7.

Although one might argue that initialization might be improved by adjusting the priors, constant priors cannot represent the expected biological variance of dwMRI signals. Indeed, even for healthy brain data, having a constant $MD_t$ prior is inadequate for representing the variability of tissue's effective diffusivities across grey and WM regions presenting different microstructural properties. Additionally, selecting reference values for the FW and tissue signals ($S_t$, $S_w$) may be inadequate to capture the spatial variation of the T2-weighted images due to bias field inhomogeneities. Moreover, it is unlikely that such complex spatial patterns can be properly captured by a trivial combination of MD and $S_o$ initializations (as done by the hybrid method).

## 5.4 Assessment of FWE-DTI specificity

Given the expected heterogeneity of healthy tissue or pathological lesions, it is important to assess the robustness of FWE-DTI algorithms in simulations with ground truth parameters that represent the spatial variation of realistic dwMRI datasets. For this, the specificity of FWE-DTI was explored on synthetic datasets generated based on the parameter estimates obtained for a representative in vivo human brain dataset using the FWE-DTI gold standard technique (NLS fitting designed particularly for multi-shell dataset (25,29)).

Although FWE-DTI techniques can provide plausible estimates from single-shell data, plausibility is not equivalent to validity. While the priors imposed by the initialization techniques seem to provide sensitivity to the high FW fractions in brain cerebral ventricles and the subarachnoid space (Figure 1), the specificity of FWE-DTI can only be ensured if changes in FW fraction can be distinguished from changes of the effective tissue diffusivity. To test for specificity, we assessed the performance of the different fitting algorithms on two different types of synthetic lesions. Our results show that when a lesion is generated for an increase of the FW fraction FWE-DTI hybrid initialization technique and consequently the RGD algorithm correctly predicts high FW volume fraction on the lesion area (Figure 5). However, when a lesion with increased $MD_t$ was simulated, single-shell techniques barely detected a change in $MD_t$ , and instead they erroneously predicted an increase in the FW fraction (Figure 6); Our results hence demonstrated that FWE-DTI specificity is not guaranteed for single-shell data.

## 5.5 Regularized descent algorithm in multi-shell data

While no advantage was observed for single-shell datasets, RGD seems to slightly improve the robustness of FW estimates. The measured FW fraction presented slightly increased precision on phantom simulations (Figure 4) and MD biases were suppressed on the MD lesion (Figure 5) when multi-shell acquisitions are considered. This result is consistent with RGD algorithms successfully converging to a more accurate solution only when multi-shell data is provided and when the FWE-DTI estimation becomes well-posed.

Interestingly, the refined FWE-DTI estimates never reached the accuracy of standard NLS FWE-DTI fitting procedures (Figures 5, 6 and 7). This might indicate that imposed spatial regularization might impede convergence to a global minimum. This contrasts with the performance of the NLS FWE-DTI algorithm which recovered the specificity of the estimates even when initialized with the single-shell hybrid initialization method. Although not shown here, single- and multi-shell RGD algorithm results were robust to changes in the learning rate and number of iterations. These observations are also in agreement with the findings of (34), where the authors noted that for data of two patients suffering from brain tumors the estimates obtained with and without the RGD algorithm were similar. In our work, we show that RGD algorithm, however, might present some benefits in terms of estimation precision.

## 5.6 Limitations and future work

The main limitation of our work is that FWE-DTI fitting routines were only tested on data and simulations reconstructed from a single healthy subject. Although not shown here, our analysis was repeated for other datasets available in DIPY showing consistent results. Since all our implementations are available in an open-source project (dipy.org), the analysis can be easily reproduced for other open access data as the Human Connectome project (42) or the Biobank (43). Although we were able to demonstrate the limitations of the RGD algorithm using simplistic simulations, it will be of interest to reproduce the results on dwMRI data of real lesions or even on physical phantoms as proposed by Farrher and colleagues (44).

In this work, tissue's non-Gaussian diffusion effects due to the presence of multiple compartments or due to the interaction between diffusing water molecules and boundaries were not considered. As we wanted to separately assess the biases introduced by the priors of the FWE-DTI initialization methods and the spatial regularization of the gradient descent algorithm, we did not include data for b-values higher than $1\ ms\ \mu m^2$. However, previous studies showed that tissue non-Gaussian effects can also introduce biases on the FWE-DTI estimates even when obtained by the well-posed multi-shell NLS algorithm (25,45). In future studies, it will be of interest to assess if these biases can further compromise the specificity of FWE-DTI estimates on both single and multi-shell datasets.

The main findings of our work point to the need of moving from single-shell acquisitions to multi-shell protocols for proper fitting of the FWE-DTI model. In addition to the information provided by multi-shell acquisitions, this study may also motivate the exploration of techniques that incorporate other sources of information. For instance, FW and tissue components may be decoupled from the signal echo time dependence since these components are expected to have different T2 relaxation properties (46). The use of this type of information might be facilitated by the rise of acquisition approaches such as ZEBRA, enabling simultaneous measurement of different diffusion-weights and echo times (47) These approaches may not only aid in better characterizing tissue compartments but also enable to assess the true fractions of different compartments.

# 6 Conclusion

In this study, we address the controversies behind the use of regularized gradient descent algorithms to fit the FWE-DTI model on single- and multi-shell in vivo and synthetic datasets. Our results show that these algorithms can provide plausible FW fraction maps on both single- and multi-shell data, due to the priors introduced upon initialization. We show, however, that based solely on these priors, FWE-DTI estimates are not able to distinguish changes in FW content from changes in $MD_t$ for single-shell data acquisitions, and thus, we stress that results from single-shell FWE-DTI in previous and future studies should be interpreted with care.

# 7 Acknowledgements

We acknowledge the Portuguese Foundation for Science and Technology (FCT - IF/00364/2013, UID/EEA/50009/2019 and UIDB/50009/2020). We also want to thank Dr Valerij G. Kiselev (Freiburg University) for insightful discussions and suggestions, and Dr Valabregue Romain (CENIR, ICM, Paris) for supplying the open source dataset used in this study.

# Figures

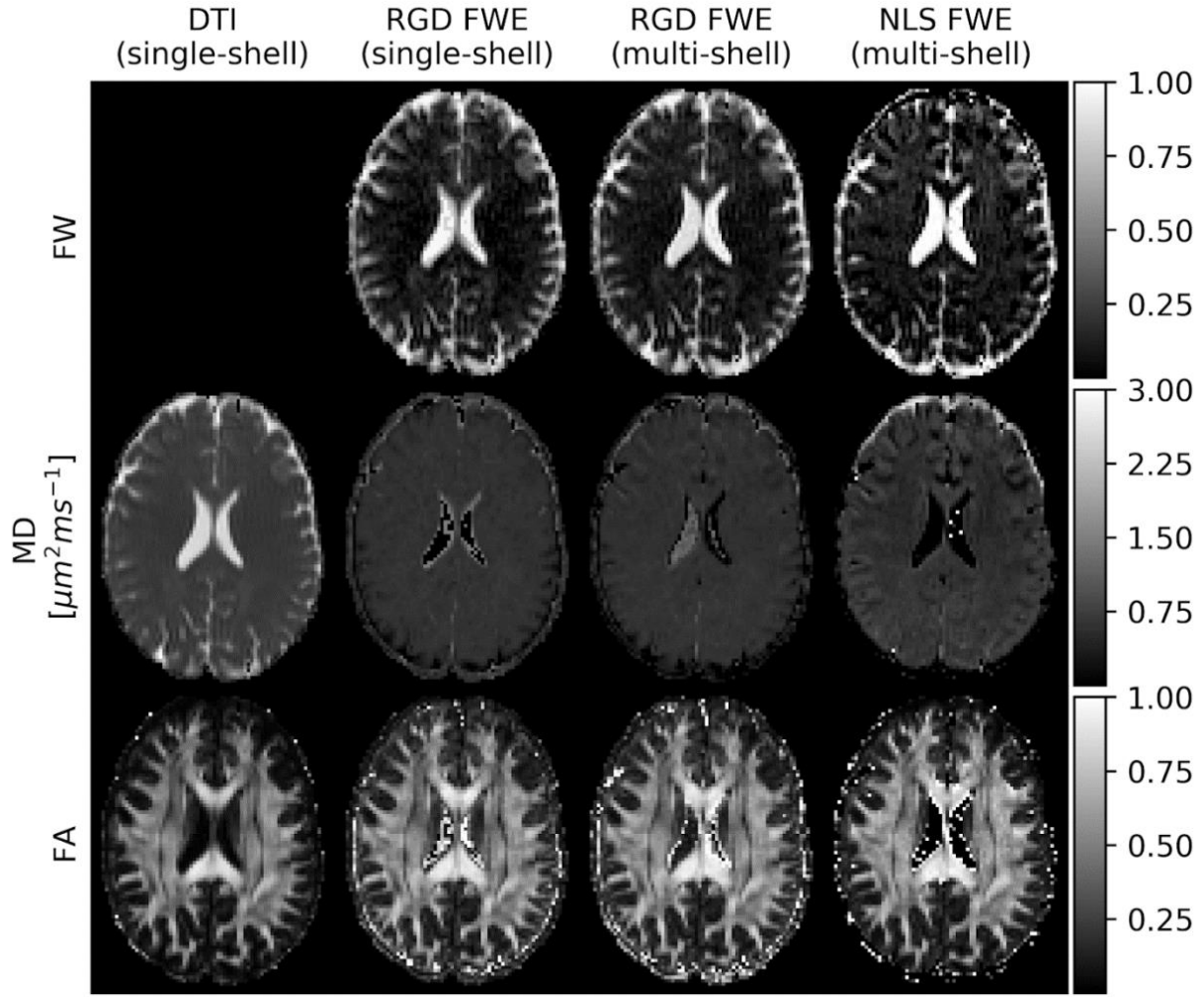

**Figure 1:** Scalar maps estimated from data acquired from a healthy volunteer (free water fraction [$f_w$], mean diffusivity [*MD*], and fractional anisotropy [*FA*] shown in the first, second and third rows, respectively), using standard DTI (first column), regularized gradient descent (RGD), FWE-DTI for single- and multi-shell data (second and third columns) and NLS FWE-DTI (fourth column). For the single-shell maps, data acquired with b-values of 1 ms µm$_{-2}$ was used; for the multi-shell maps, all b-values up to 1 ms µm$_{-2}$ were used.

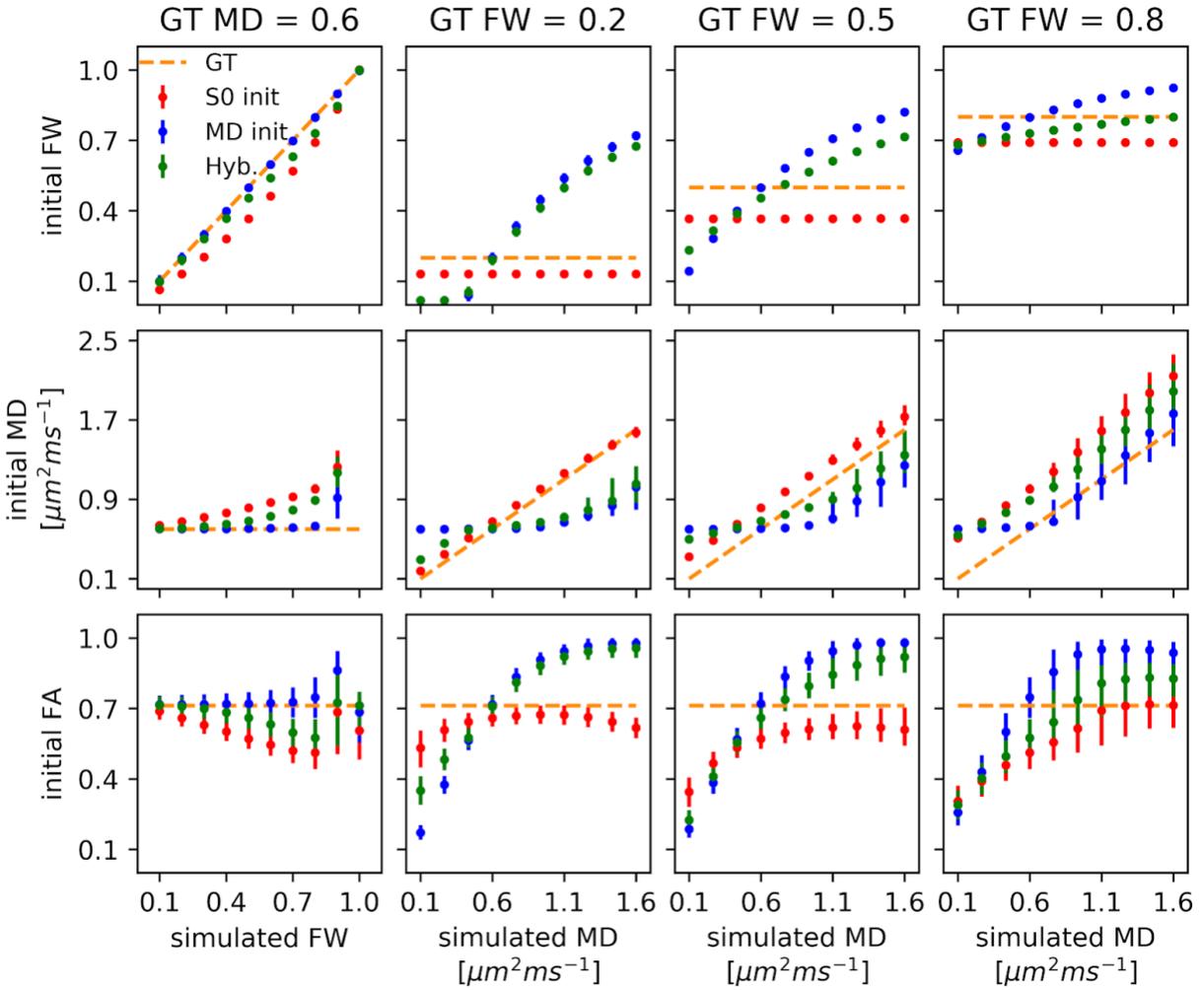

**Figure 2:** Median and interquartile ranges of the initial $f_w$, $MD_t$ and $FA$ (top, middle and bottom rows, respectively) applied to single-voxel signals over 10,000 repeats (100 $D_t$ directions × 100 noise instances). These values are plotted as a function of $f_w$ ground truth values (from 0 to 1) on the left column (for a fixed ground truth $MD_t$=0.6 μm²ms⁻¹) and plotted as a function of $MD_t$ ground truth values (from 0.1 to 1.6 $\mu m^2 ms^{-1}$) on the second to fourth columns (for fixed $f_w$ ground truth values of 0.2, 0.5 and 0.8, respectively). On each panel, ground truth values are represented by the orange lines, $S_0$ initialization estimates by the red markers, $MD_t$ initialization estimates by the blue markers, while hybrid initialization corresponds to the green markers.

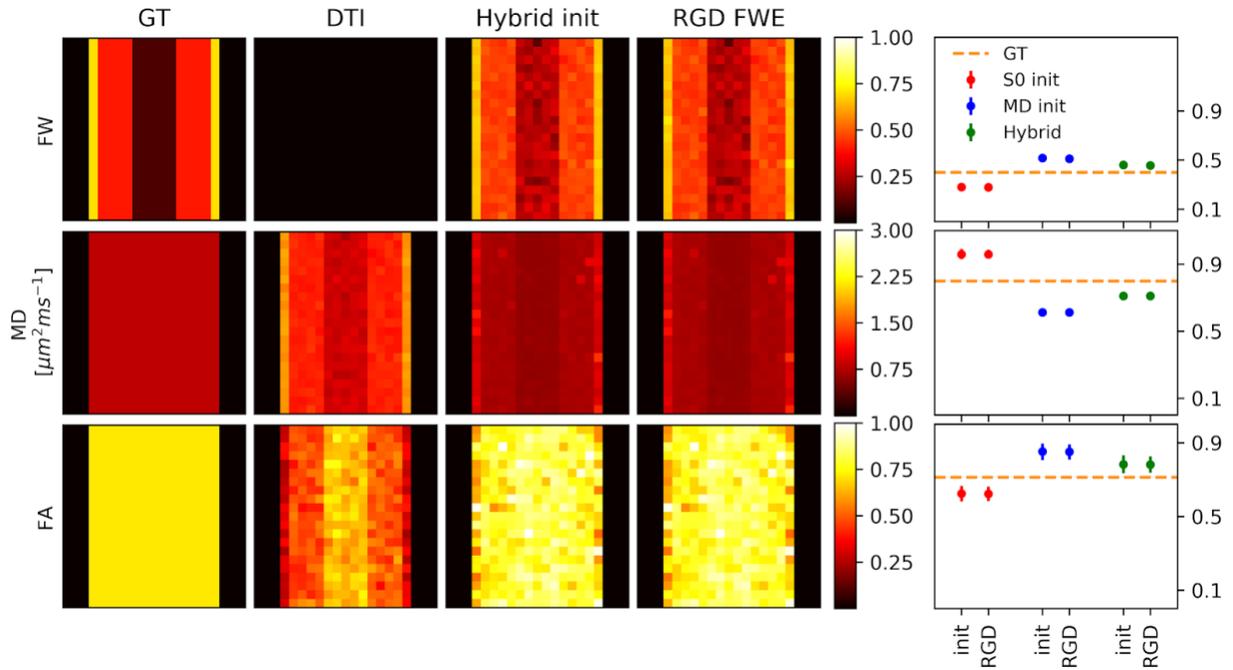

**Figure 3:** Results for the single-shell dataset of the multi-voxel phantom. On the left side of the figure, the $f_w$, *FA* and *MD* parametric map estimates are shown for the ground truth values (first column), standard DTI (second column), FWE-DTI hybrid initialization (third column), and FWE-DTI regularized gradient descent (RGD) algorithm initialized using the hybrid method (fourth column). On the right side, the corresponding distributions are plotted for the estimates before ('init') and after ('est') applying the RGD algorithm and for the three initialization methods - $S_0$, $MD_t$ and hybrid initializations (red, blue and green, respectively) compared to the GT (orange). To enable an easier interpretation, the median and interquartile rates were computed inside the region with intermediate GT $f_w$ value of 0.4 (similar conclusions could be drawn for the other GT $f_w$ values - results not shown).

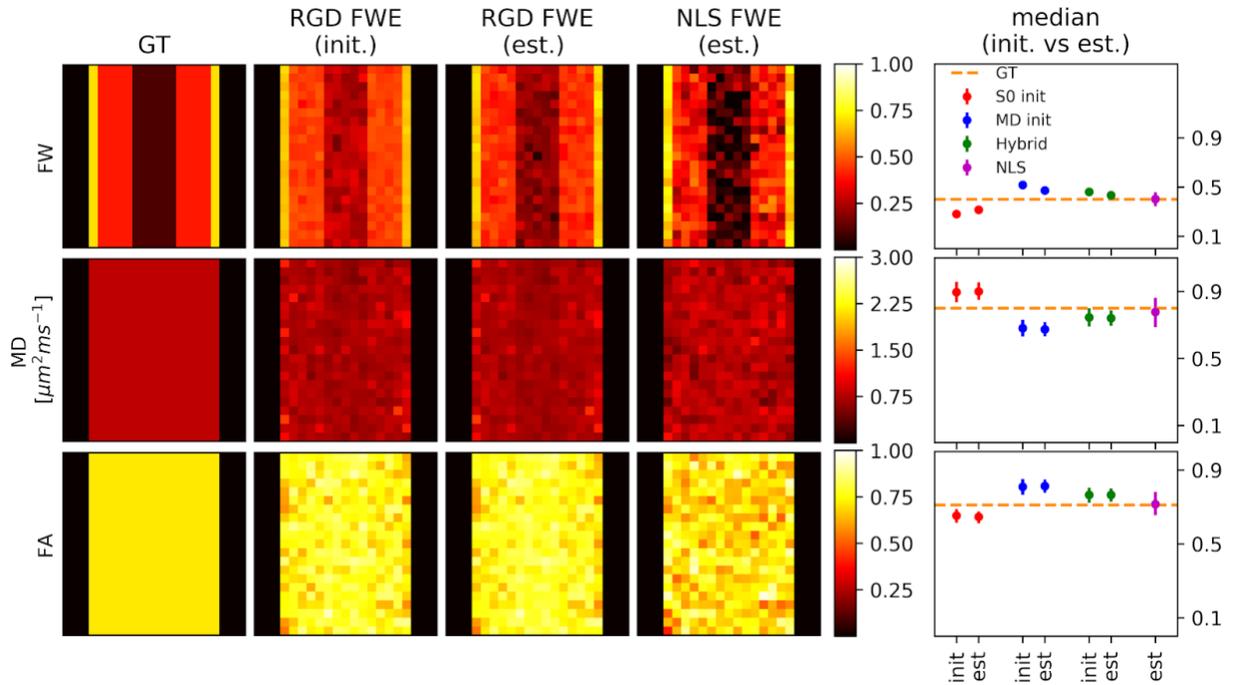

**Figure 4:** Results for the multi-shell dataset of the multi-voxel phantom. On the left side of the figure, the parametric maps of fw, FA and MD estimates are shown for the ground truth values (first column), FWE-DTI hybrid initialization (second column), FWE-DTI regularized descent algorithm (RGD) initialized using the hybrid method (third column), and the NLS FWE-DTI algorithm (fourth column). On the right side of the figure, the corresponding distributions are plotted for the estimates before ('init') and after ('est') applying the RGD algorithm and for the three initialization methods - S0, MDt and hybrid initialization (red, blue and green, respectively) compared to the GT (orange). To enable an easier interpretation, the median and interquartile rates were computed inside the region with intermediate GT fw value of 0.4 (similar conclusions could be drawn for the other GT fw values - results not shown).

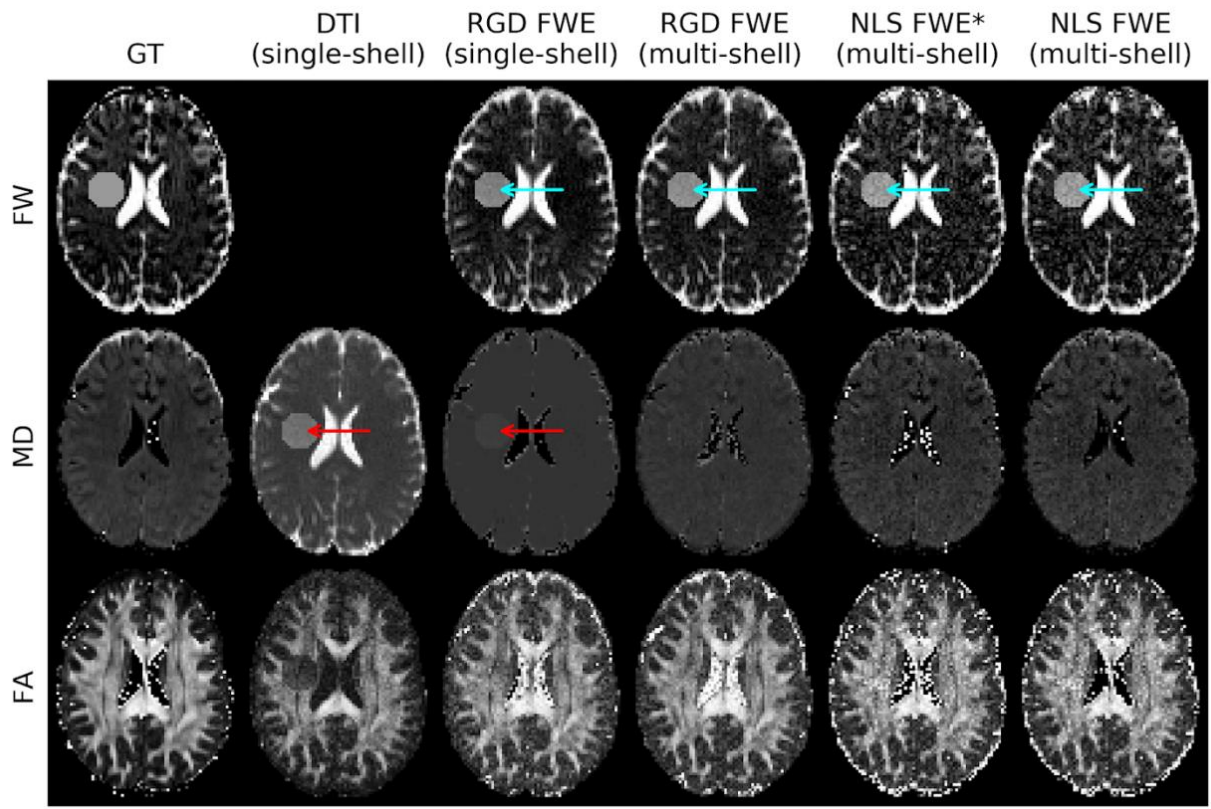

**Figure 5:** Scalar maps estimated from in vivo data after introducing a simulated FW lesion. The ground truth (GT) is represented on the first column, while the remaining columns show the estimates obtained using standard DTI (second column), RGD FWE-DTI for single- and multi-shell data (third and fourth columns) and NLS FWE-DTI (fifth column). The NLS FWE-DTI* (sixth column) shares the same initialization method used in the RGD routine (hybrid approach). The single-shell data was simulated along 32 directions with b=1 ms µm$_{-2}$ (in addition to six b-value=0 images); the multi-shell data was simulated with b-values of 0.5 ms µm$_{-2}$ and 1 ms µm$_{-2}$ (32 directions each, in addition to six b-value=0 images).

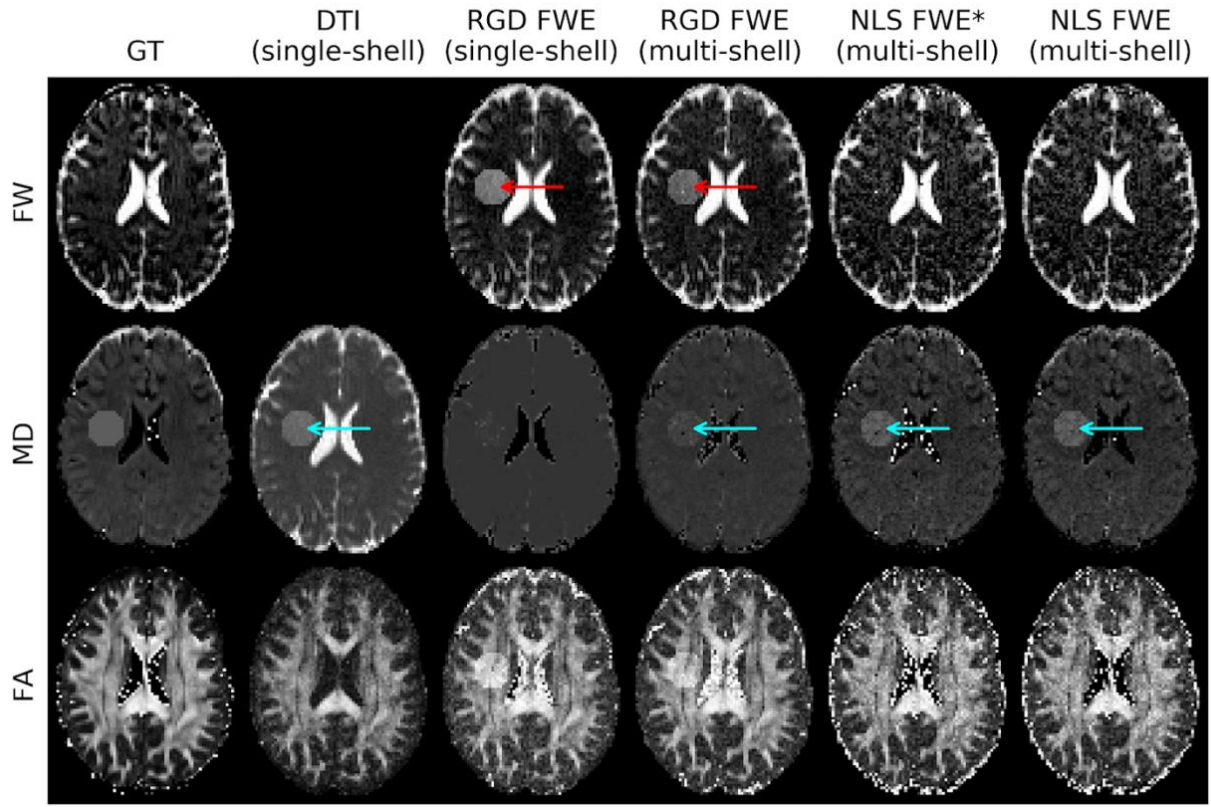

**Figure 6:** Scalar maps estimated from in vivo data after introducing a simulated MD lesion. The ground truth (GT) is represented on the first column, while the remaining columns show the estimates obtained using standard DTI (second column), RGD FWE-DTI for single- and multi-shell data (third and fourth columns) and NLS FWE-DTI (fifth column). The NLS FWE-DTI* (sixth column) shares the same initialization method used in the RGD routine (hybrid approach). The single-shell data was simulated along 32 directions with b=1 ms µm$_{-2}$ (in addition to six b-value=0 images); the multi-shell data was simulated with b-values of 0.5 ms µm$_{-2}$ and 1 ms µm$_{-2}$ (32 directions each, in addition to six b-value=0 images).

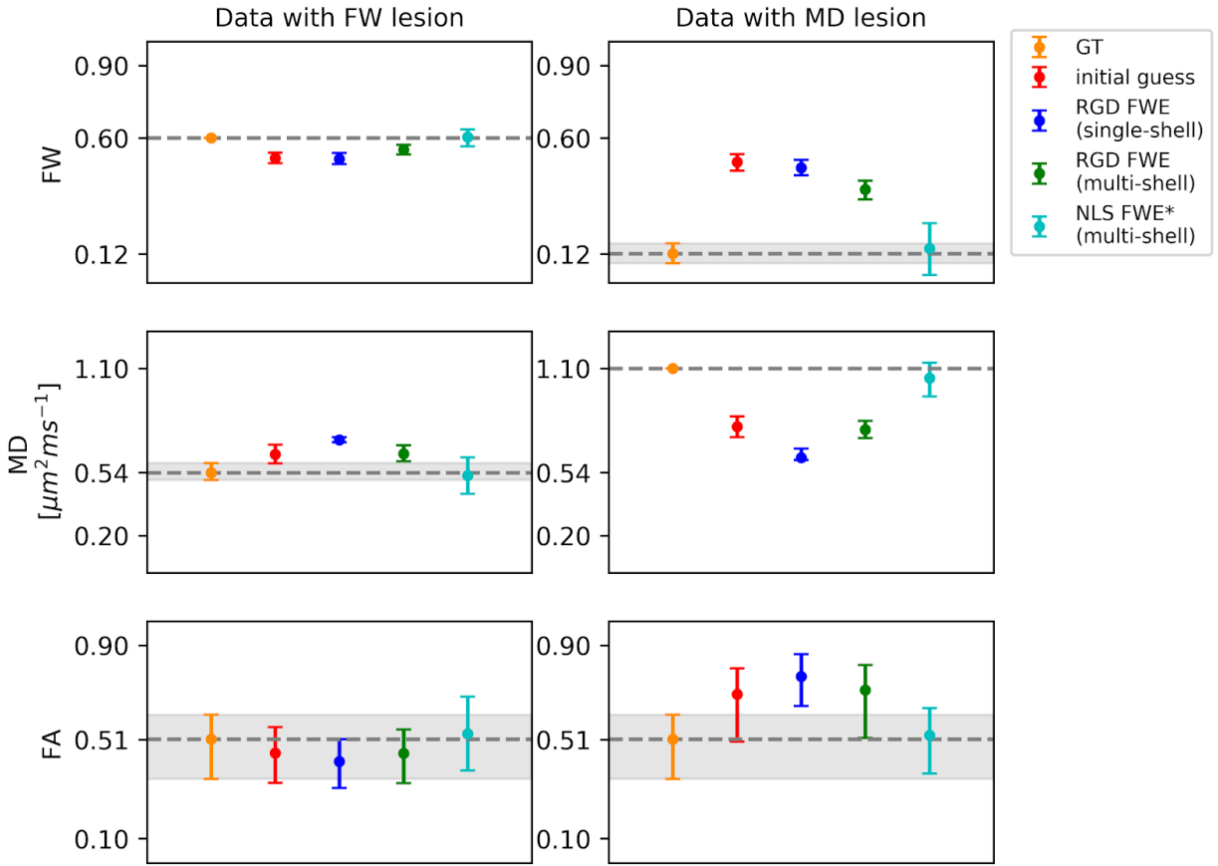

**Figure 7:** Distribution of the $f_w$, $MD_t$ and $FA$ estimates (first, second and third rows) extracted from the FW lesion (left panels) and MD lesion (right panels). The markers correspond to the median values while the bars represent the interquartile range computed inside the lesion. On each panel estimates from the initial hybrid method, RGD single- and multi-shell methods and modified NLS fitting routines are plotted in red, blue, green, and cyan respectively and compared to the ground truth (GT) (orange). To facilitate comparisons, a dashed grey line indicates the median GT value, with the shaded areas representing its interquartile range.